\documentclass[sigchi-a]{acmart}

\AtBeginDocument{%
  \providecommand\BibTeX{{%
    \normalfont B\kern-0.5em{\scshape i\kern-0.25em b}\kern-0.8em\TeX}}}

\copyrightyear{2019}
\acmYear{2019}

\begin{document}

\title{Chat-Bot-Kit: A web-based tool to simulate text-based interactions between humans and with computers}

\author{Kyoko Sugisaki}
\email{sugisaki@ds.uzh.ch}
\affiliation{%
  \institution{German department, University of Zurich}
  \streetaddress{Sch\"{o}nberggasse 9}
  \city{Zurich}
  \state{Switzerland}
  \postcode{8001}
}


\begin{abstract}
In this paper, we describe Chat-Bot-Kit, a web-based tool for text-based chats that we designed for research purposes in computer-mediated communication (CMC). Chat-Bot-Kit enables to carry out language studies on text-based \textit{real-time} chats for the purpose of research: The generated messages are structured with language performance data such as pause and speed of keyboard-handling and the movement of the mouse. The tool provides two modes of chat communications -- quasi-synchron and synchron modes-- and various typing indicators. The tool is also designed to be used in wizard-of-oz studies in Human-Computer Interaction (HCI) and for the evaluation of chatbots (dialogue systems) in Natural Language Processing (NLP).
\end{abstract}

%


\maketitle

\section{Chat tool}
Text messaging is a very popular form of communication, and apps like Facebook Messenger, SnapChat, WeChat, or WhatsApp belong to the most installed apps worldwide. Text messaging is, in recent years, a hotbet for technical and user experience innovations: Small changes of rules and design lead to completely different usage patterns, usage scenarios and overall user experiences. As an example, Snapchat lets the messages disappear  a short time after the are consumed/read. Just by doing so, messaging becomes a more ephemeral form of communication that does not leave traces or tangible artefacts behind and is providing a safer and trusted space to exchange more sensitive and private topics. WhatsApp lets users see the status of messages, i.e. whether they are transmitted successfully to the recipient or read, and it shows the online activity of others, which renders the interaction again more closely to face-to-face communication, compared to classical text messaging services such as SMS.

\vspace{3mm}
\noindent 
In recent years, major tech companies have also started to offer chatbot platforms that allow businesses to automate conversations with consumers and reach them where they spend a lot of their time (i.e. within popular messaging apps). On Facebook Messenger alone, more than 300.000 chatbots have been deployed by Mid 2018. Many industry observers claim that chatbots development will replace app development. At the same time, consumers struggle with the interactions \cite{Brandtzaeg2018,Folstad2017}. At the current state of the technology, the text-based chat communication with computers is needed to be better understood, more actively designed and more frequently tested to make them more usable and valuable for users, let alone feel more natural. A common way to simulate conversations with chatbots in a cheap, versatile way without actually implementing them is to conduct so called Wizard of Oz studies (woz), in which a researcher pretends to be a computer during a conversation with a human test subject who is briefed to talk with a computer. Such studies are common in the field of CMC, Natural Language Processing (NLP) and Human-Computer-Interaction.

\begin{figure}[t]
\hfill
\includegraphics[width=7cm,height=6cm]{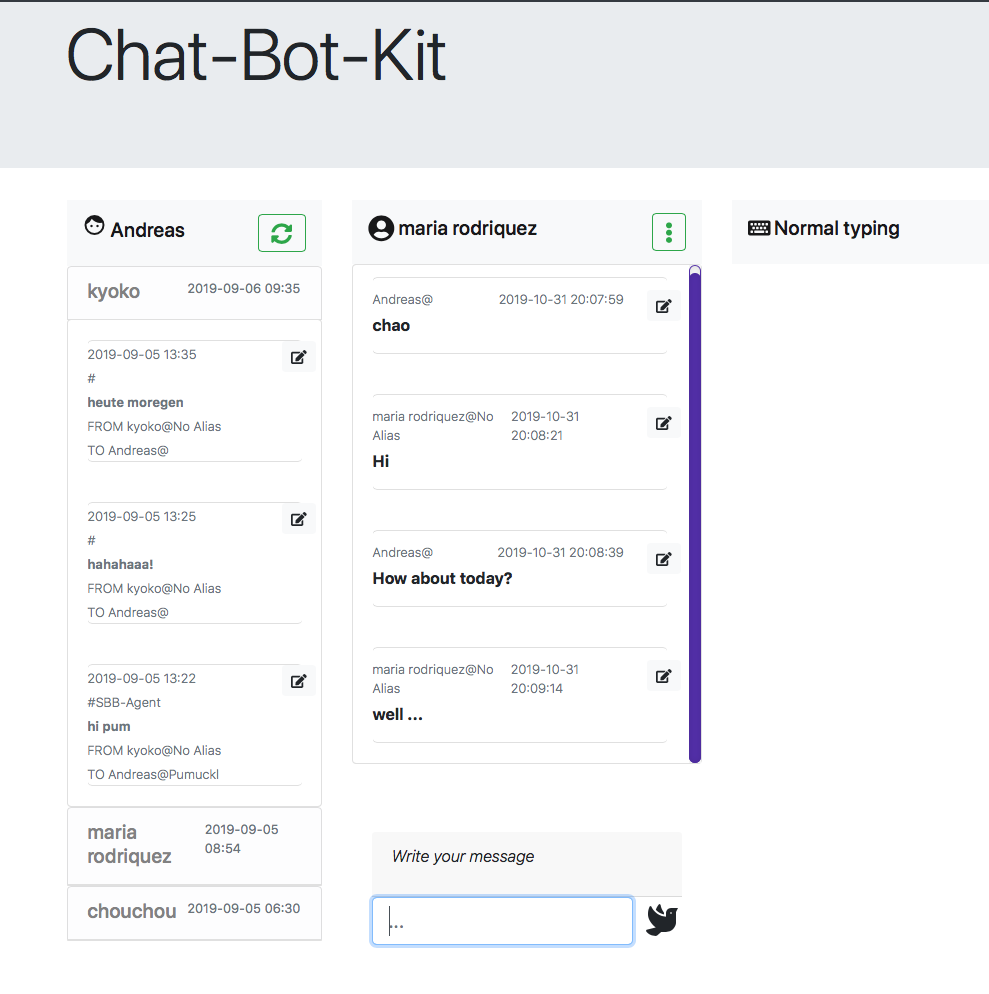}
\hfill
\includegraphics[width=7cm,height=6cm]{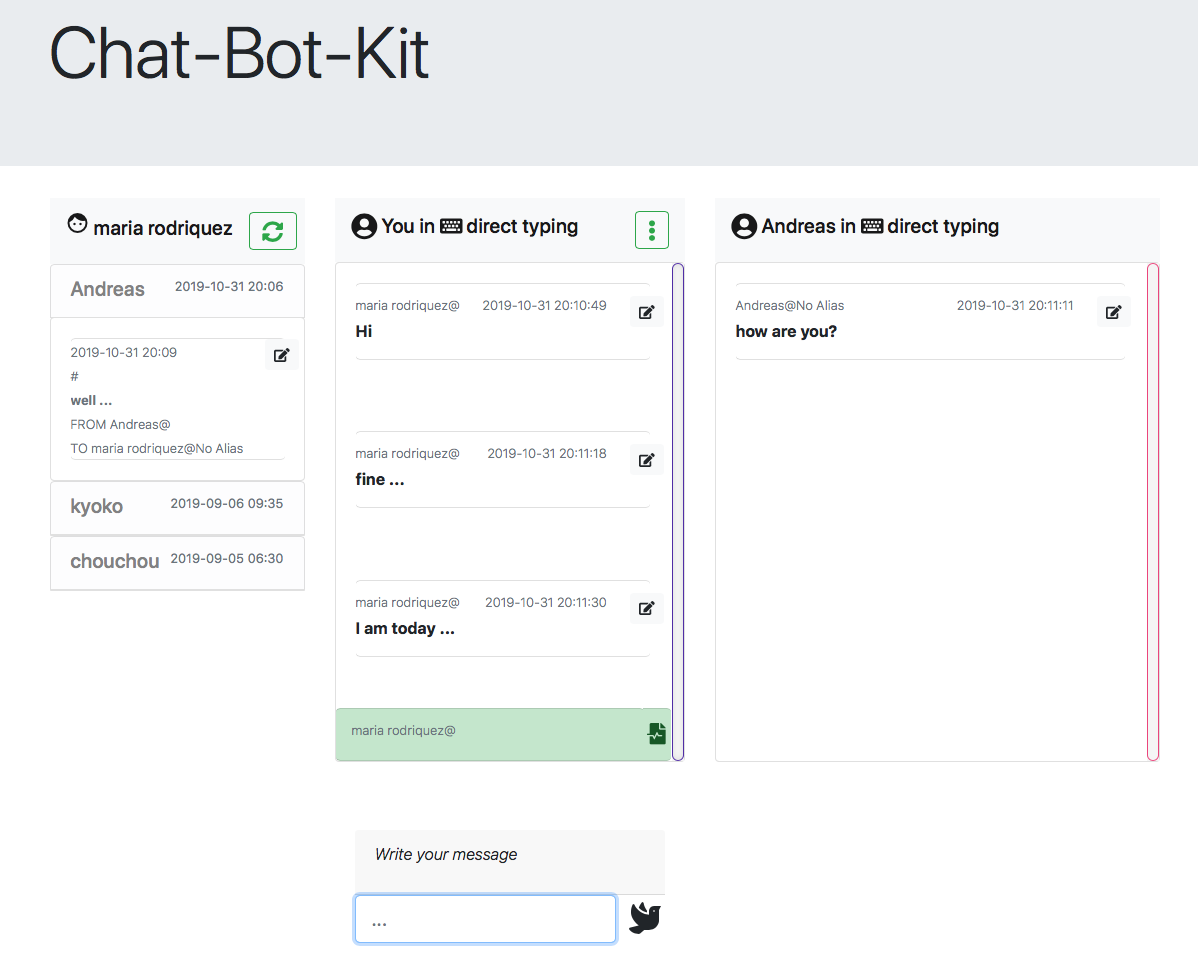}
\hfill
\caption{Chat-Bot-Kit}
\label{fig:cbk}
\end{figure}

\vspace{3mm}
\noindent
There are a number of tools with which one could simulate chats and WoZ-studies. One could use a tool like Skype, for example. There are also some commercial chat tools that provide a log file that is useful for the purpose of research (e.g., \cite{Thies2017}). However, there are some missing features for the purpose of research. 

Chat-Bot-Kit, a web-based chat tool (Fig. 1) that we present in this paper has been developed based on the following requirements for research in Computer-mediated communication (CMC), Human-Computer Interaction (HCI) and Natural Language Processing (NLP). 
\begin{itemize}

	\item The submitted messages can be exported into a structured data format (Microsoft Office Excel) for the purpose of further analysis on the chat communication.  
	\item The language performance data is embedded into the structure of messages. The tool automatically measures the pause, speed, rhythm of keyboard stokes and the movement of the mouse, next to the time stamp of the message submission. The temporal feature of the chat interaction is of relevance in CMC (cf. \cite{Jones2013})  
	\item A user can have more than one name and role: this is considered to be used in HCI. In case of a wizard-of-oz simulation study \cite{Li2007,Dow2005,Dahlback1993}, the identity of chat constructors (for example, an agent of an insurance company) plays an important role for the credibility of wizards as machine.
	\item The submitted messages can be edited, rated and commented directly in the user interface and the data is integrated into the output file. The feature is considered for HCI and NLP studies in which the designer or developer of a chatbot wants to test the system and carry out the user evaluation on the level of messages.
	\item The tool is scalable: the system is based on AngularJS components that are easy to be extended to a wizard assistant tool or a chatbot
	\item There are several options of the chat communication with regard to the turn-taking. 
	\begin{itemize}
	\item Typing indicator: the typing indicator \cite{Gnewuch2018} displays the typing behaviour of the communication partner. This feature affects the turn-taking, as it is a type of mutual monitoring between communicators. In our tool, the typing indicator can be configured by the communicators and study leaders.
	\item Quasi-synchronous and synonymous mode: In the CMC, the text-chat communication is currently investigated in the quasi-synchronous mode. The quasi-synchronous mode has dominated text-based chats (such as WhatsApp or Skype), thus has been investigated for a long time (e.g., \cite{Garcia1999,Herring1999}). While quasi-synchronous written communication is regarded as one-way transmission of turns and the on-going typing is only visible to the writer, but not to the other participants, synchronous written communication is characterised in two-way transmission where the participants transmit their typing in a keystroke-by-keystoke to their own window, instead of into one window for all interlocutors in the quasi-synchronous one. In the past, there were frameworks such UNIX talk or VAX phone for the synchronous one (cf.\ \cite{Herring1999}). Google Wave was also experimenting with synchronous modes. The synchronicity of the communication is one crucial difference in face-to-face/phones and CMS that affects the interaction deeply, in particular in turn-taking  \citep{Garcia1999,Herring1999} and repair \citep{Jacobs2013}. The tool provides these two modes of CMC, ready to be used in studies.
	\end{itemize} 
\end{itemize}

\section{Conclusion}
In this paper, we presented our web-based chat tool designed for the research in Computer-mediated communication (CMC), Human-Computer Interaction (HCI) and Natural Language Processing (NLP). 
In future work, we plan to extend several wizard assistance methods for wizard-of-oz studies in HCI that include machine learning and allow to iteratively train models during a study.

\bibliographystyle{ACM-Reference-Format}
\bibliography{sample-base}


\end{document}